\documentclass[final]{IEEEtran}
\usepackage{amsmath, amssymb}
\usepackage{etoolbox}
\usepackage{graphicx}
\usepackage{subfigure}
\usepackage[noadjust]{cite}
\newtheorem{theorem}{Theorem}
\usepackage{color}

\title{Smart Transmission Network Emergency Control}

\author{Thanh Long Vu,~\IEEEmembership{Member,~IEEE,} Hsiao-Dong Chiang,~\IEEEmembership{FIEEE,} and~Konstantin~Turitsyn,~\IEEEmembership{Member,~IEEE}
\thanks{Thanh Long Vu and Konstantin Turitsyn are with the Department of Mechanical Engineering, Massachusetts Institute of Technology, Cambridge, MA, 02139 USA, email: \{longvu, turitsyn\}@mit.edu. Hsiao-Dong Chiang is with School of Electrical and Computer Engineering, Cornell University,  Ithaca, NY, USA, email: chiang@ece.cornell.edu.
}}

\begin{document}

\maketitle
\begin{abstract}
Power systems normally operate at their stable operating conditions where the power supply and demand are balanced. In emergency situations, the operators proceed to cut a suitable amount of loads to rebalance the supply-demand and hopefully stabilize the system. This traditional emergency control scheme results in interrupted service with severely economic damages to customers. In order to provide seamless electricity service to customers, this paper proposes a viable alternative for traditional remedial controls of power grids by exploiting the plentiful transmission facilities. In particular, we consider two emergency control schemes involving adjustment of the susceptance of a number of selected transmission lines to drive either fault-on dynamics or post-fault dynamics, and thereby stabilize the system under emergency situations. The corresponding emergency control problems will be formulated and partly solved in some specific cases. Simple numerical simulation will be used to illustrate the concept of this paper.
\end{abstract}

\maketitle

\section{Introduction}

High penetration of intermittent renewables, large volume of power storage and EVs, and increasing load demand are pushing the aging US power grid to its physical limits. Consequently, the stressed system is especially vulnerable to large disturbances. The current emergency controls are based on remedial actions \cite{119276,6965423}, special protection systems (SPS) \cite{SPS,982194} and load shedding \cite{141798, 1461637} to quickly rebalance power and hopefully stabilize the system. However, some of these emergency actions rely on interrupting electricity service to customers. The unexpected service loss is extremely harmful to customers since it may lead to enormously high economic damage. In addition, the protective devices are usually only effective for individual elements, but less effective in preventing the grid from collapse. For example, the recent major blackouts witness the inability of operators in preventing the grid from cascading failures \cite{2003blackout}, regardless the good performance of  the individual protective devices. The underlying reason is the lack of coordination among protective devices and the difference in their timescales, which together make them incapable to maintain the grid stability. This calls for system-level solutions for emergency control of power grids.

On the other hand we note that the US electric infrastructure currently contains approximately 642,000 miles of high-voltage transmission lines and almost 6.3 million miles of electricity distribution lines, incorporated with plentiful control equipments such as FACTS devices. The key idea of this paper is to extract more value out of the existing transmission and distribution facilities to stabilize power systems under emergency situations. Particularly, we propose to use FACTS devices to adjust susceptances of a number of selected transmission lines to drive the fault-on dynamics or post-fault dynamics and thereby stabilize the power system. 

To this end, this paper will formulate  two emergency transmission control problems: one involves controlling the fault-on dynamics to maintain the transient stability of the system following a line tripping and the other involves controlling the post-fault dynamics to stabilize a given fault-cleared state that is otherwise possibly unstable. As a first step, we will sketch ways to solve these emergency control problems in specific cases by applying our recently introduced quadratic Lyapunov function-based transient stability certificate \cite{VuTuritsyn:2015TAC,VuTuritsyn:2014}. In particular, we present sufficient conditions for the susceptance values such that when applied to the fault-on dynamics or post-fault dynamics, the fault-cleared state will stay inside the region of attraction of the post-fault equilibrium point. Remarkably, the sufficient conditions are formulated as a set of linear matrix inequalities (LMIs), which can be solved quickly by advanced network structure-exploiting convex optimization solvers \cite{Jabr2012}. As such, the proposed emergency control schemes can be scalable to large-scale power systems.


\section{Network Model and Emergency Control Problems}
\label{sec.model}

\subsection{Network Model}

In this paper we consider the standard structure-preserving model to describe components and dynamics in
power systems \cite{bergen1981structure}. This model naturally
incorporates the dynamics of generators' rotor angle as well as response of
load power output to frequency deviation. 
Mathematically, the grid is described by an undirected graph
$\mathcal{A}(\mathcal{N},\mathcal{E}),$ where
$\mathcal{N}=\{1,2,\dots,|\mathcal{N}|\}$ is the set of buses and
$\mathcal{E} \subseteq \mathcal{N} \times \mathcal{N}$ is the set
of transmission lines connecting those buses. Here, $|A|$ denotes
the number of elements in the set $A.$ The sets of generator buses
and load buses are denoted by $\mathcal{G}$ and $\mathcal{L}$. We assume that the grid is lossless with
constant voltage magnitudes $V_k, k\in \mathcal{N},$ and the
reactive powers are ignored. Then, the structure-preserving model of the system is given by:
\begin{subequations}
\label{eq.structure-preserving}
\begin{align}
\label{eq.structure-preserving1}
 m_k \ddot{\delta_k} + d_k \dot{\delta_k} + \sum_{j \in
  \mathcal{N}_k} a_{kj} \sin(\delta_k-\delta_j) = &P_{m_k},  k \in \mathcal{G},  \\
  \label{eq.structure-preserving2}
  d_k \dot{\delta_k} + \sum_{j \in
  \mathcal{N}_k} a_{kj} \sin(\delta_k-\delta_j) = &-P^0_{d_k},  k \in \mathcal{L},
\end{align}
\end{subequations}
where, the equations \eqref{eq.structure-preserving1} represent
the dynamics at generator buses and the equations
\eqref{eq.structure-preserving2} the dynamics at  load buses.
In these equations, with $k \in \mathcal{G}$ then
 $m_k>0$ is the dimensionless moment of inertia of the
generator, $d_k>0$ is the term representing primary frequency
controller action on the governor, $P_{m_k}$ is the input shaft power producing the mechanical torque acting on the rotor, and $P_{e_k}$
is the effective dimensionless electrical power output of the
$k^{th}$ generator. With $k \in \mathcal{L}$ then $d_k>0$ is  the constant frequency coefficient of load and 
$P^0_{d_k}$ is the nominal load.
Here, $a_{kj}=V_kV_jB_{kj},$ where $B_{kj}$ is the
(normalized)  susceptance of the transmission line $\{k,j\}$ connecting the $k^{th}$ bus and $j^{th}$ bus,
$\mathcal{N}_k$ is the set of neighboring buses of the $k^{th}$
bus.   Note that the system described by equations \eqref{eq.structure-preserving}
has many stationary points characterized
by the angle differences $\delta_{kj}^*=\delta_k^*-\delta_j^*$
that solve the following system of power-flow like equations:
\begin{align}
  \label{eq.SEP}
  \sum_{j \in
  \mathcal{N}_k} a_{kj} \sin(\delta_{kj}^*) =P_{k}, k \in \mathcal{N},
\end{align}
where $P_k=P_{m_k}, k \in \mathcal{G},$ and $P_k=-P^0_{d_k}, k \in
\mathcal{L}.$



\subsection{Emergency Control Problems}
\label{sec.formulation}

In normal conditions, a power grid operates at a stable equilibrium point of
the pre-fault dynamics. Under emergency situations caused by a fault,
the system evolves according to the fault-on dynamics laws and moves away
from the pre-fault equilibrium point $\delta^*_{pre}$ to a fault-cleared state $\delta_0=\delta_{fault-cleared}$ at the clearing time $\tau_{clearing}$. 
Our emergency control objective is to make sure that post-fault dynamics is stable. i.e. there exists a stable post-fault equilibrium point $\delta^*_{post}$ whose region of attraction (i.e. stability region) contains the fault-cleared state.
We consider two control schemes to achieve this objective: (i) adjusting the fault-on dynamics and (ii) adjusting the post-fault dynamics. Both schemes are based on changing the susceptance of some selected transmission lines through FACTS devices. 

For the fault-on dynamics controlling, we consider the case when a system is being at the stable equilibrium point and then one line is tripped. We will adjust susceptances of some selected transmission lines such that the fault-on trajectory does not deviate too much from the equilibrium point. Hence, when the line is reclosed at the clearing time $\tau_{clearing},$ the fault-cleared state will still stay inside the region of attraction of the equilibrium point, i.e. the post-fault dynamics is stable.

\begin{itemize}
\item [(\textbf{P1})] \textbf{Emergency Control on Fault-on Dynamics} \emph{Given a fault causing tripping of line $\{u,v\}$ and the post-fault equilibrium point $\delta^*_{post},$ determine the feasible values for susceptances of selected transmission lines such that the fault-cleared state $\delta_0(\tau_{clearing})$ at the given clearing time $\tau_{clearing}$ is always inside the nominal region of attraction of the post-fault equilibrium point $\delta^*_{post}$.}
\end{itemize}

\begin{figure}[t!]
\centering
\includegraphics[width = 3.2in]{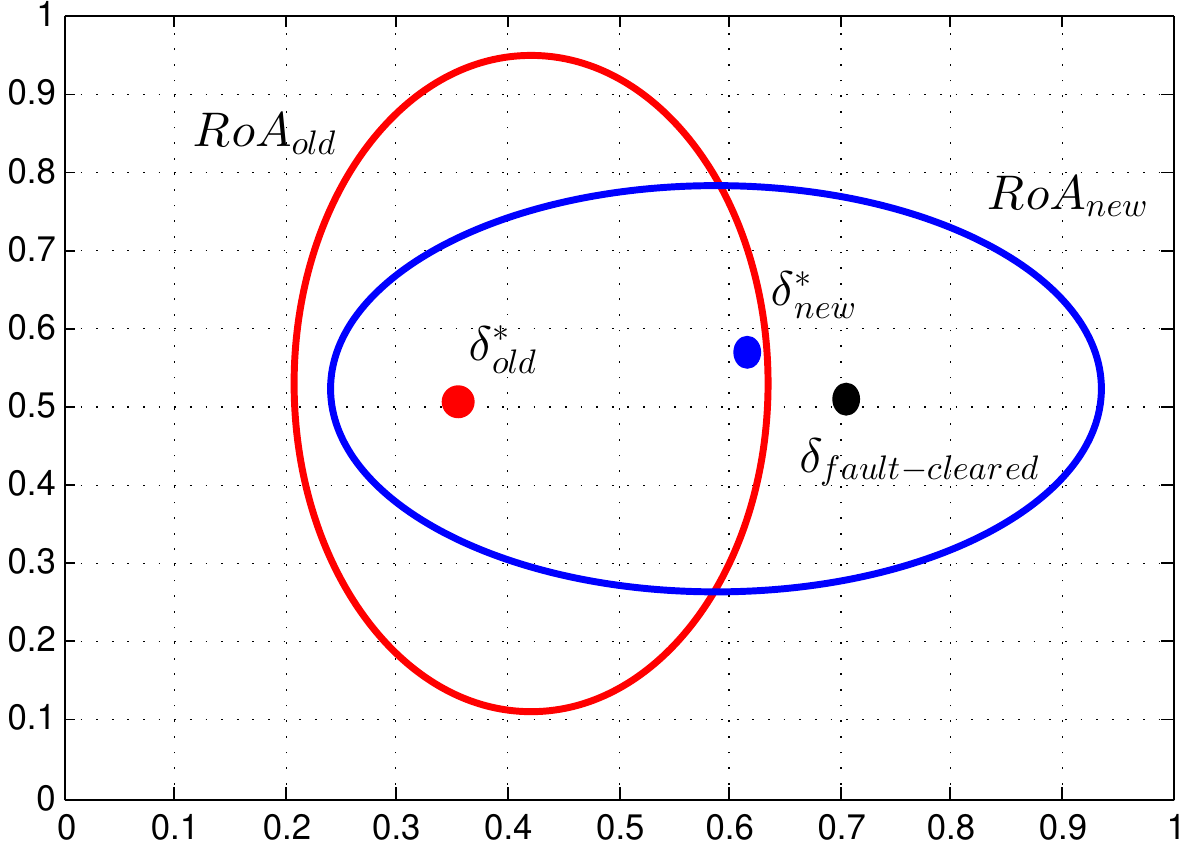}
\caption{Stability-driven smart transmission control: the fault-cleared state $\delta_{fault-cleared}$ is made stable by changing post-fault dynamics through adjusting susceptances of selected transmission lines}
\label{fig.EmergencyControl_Idea}
\end{figure}

For the post-fault dynamics controlling, we are interested in the case, showed in Fig. \ref{fig.EmergencyControl_Idea}, when a given post-fault dynamics is possibly unstable as the fault-cleared state $\delta_0$ may stay outside the region of attraction of the equilibrium point $\delta^*_{old}$. Yet by changing the susceptance of a number of transmission lines, we can obtain a new post-fault dynamics with new equilibrium point $\delta^*_{new}$ whose region of attraction contains  the fault-cleared state $\delta_0$, and therefore the new post-fault dynamics is stable.  
\begin{itemize}
\item [(\textbf{P2})] \textbf{Emergency Control on Post-fault Dynamics:} \emph{Given a fault-cleared state $\delta_0,$ determine the feasible values for susceptances of selected transmission lines such that the fault-cleared state $\delta_0$ is inside the region of attraction of the new post-fault equilibrium point $\delta^*_{new}$.}
\end{itemize}

\section{Quadratic Lyapunov Function-based Transient Stability Certificate}
\label{sec.certificates}

In this section, we recall our recently introduced quadratic Lyapunov function-based transient stability certificate \cite{VuTuritsyn:2015TAC} for power systems with the post-fault equilibrium point $\delta^*$, which will be instrumental to designing emergency controls in the next sections.  To this end, we
separate the nonlinear couplings and the linear terminal system in
\eqref{eq.structure-preserving}. Consider the state vector $x =
[x_1,x_2,x_3]^T,$ which is composed of the vector of generator's
angle deviations from equilibrium $x_1 = [\delta_1 -
\delta_1^*,\dots, \delta_{|\mathcal{G}|} -
\delta_{|\mathcal{G}|}^*]^T$, their angular velocities $x_2 =
[\dot\delta_1,\dots,\dot\delta_{|\mathcal{G}|}]^T$, and vector of
load buses' angle deviation from equilibrium
$x_3=[\delta_{{|\mathcal{G}|}+1}-\delta_{{|\mathcal{G}|}+1}^*,\dots,\delta_{|\mathcal{N}|}-\delta_{|\mathcal{N}|}^*]^T$.
Let the matrix $C$ such that 
$Cx=[(\delta_{kj}-\delta_{kj}^*)_{\{k,j\}\in\mathcal{E}}]^T.$
Consider the vector of nonlinear interactions $F$ in the simple trigonometric form: $
F(Cx)=[(\sin\delta_{kj}-\sin\delta^*_{kj})_{\{k,j\}\in\mathcal{E}}]^T.$ Denote
the matrices of moment of inertia, frequency controller  action on governor, and  frequency coefficient of load
as $M_1=\emph{\emph{diag}}(m_1,\dots,m_{|\mathcal{G}|}), D_1=\emph{\emph{diag}}(d_1,\dots,d_{|\mathcal{G}|})$ and $M=\emph{\emph{diag}}(m_1,\dots,m_{|\mathcal{G}|},d_{|\mathcal{G}|+1},\dots,d_{|\mathcal{N}|}).$
Then, the power system \eqref{eq.structure-preserving} can be then expressed as follows:

\begin{equation}\label{eq.Bilinear}
 \dot x = A x - B F(C x),
\end{equation}
with the matrices $A,B$ given by the following expression:
\begin{align*}
A=\left[
        \begin{array}{ccccc}
          O_{m \times m} \qquad & I_{m \times m}  \qquad & O_{m \times n-m}\\
          O_{m \times m} \qquad & -M_1^{-1}D_1 \qquad & O_{m \times n-m} \\
          O_{n-m \times m} \qquad &O_{n-m \times m} \qquad & O_{n-m \times n-m}
        \end{array}
      \right],
\end{align*}
and $B= \left[
        \begin{array}{ccccc}
          O_{m \times |\mathcal{E}|}; \quad
          S_1M^{-1}E^TS; \quad
          S_2M^{-1}E^TS
        \end{array}
      \right],$
      where $n=|\mathcal{N}|, m=|\mathcal{G}|,$ $S=\emph{\emph{diag}}(a_{kj})_{\{k,j\}\in \mathcal{E}}, $
$S_1=[I_{m\times m}\quad O_{m\times n-m}], S_2=[O_{n-m\times m} \quad I_{n-m\times n-m}].$

The construction of quadratic Lyapunov function is based on the bounding of the nonlinear term $F$ by linear functions of the angular differences. Particularly, 
we observe that for all values of
$\delta_{kj} = \delta_k - \delta_j$  staying inside the polytope $\mathcal{P}$ defined by the inequalities $|\delta_{kj}|  \le\pi/2,$ we have:
\begin{align*}
g(\delta_{kj}-\delta_{kj}^*)^2 \le (\delta_{kj}-\delta_{kj}^*)(\sin\delta_{kj} - \sin\delta_{kj}^*) \le (\delta_{kj}-\delta_{kj}^*)^2
\end{align*}
where
 $g=\min_{\{k,j\}\in \mathcal{E}}{(1-\sin|\delta_{kj}^*|)}/{(\pi/2-|\delta_{kj}^*|)}.$ With $\delta^*$ staying inside the polytope $\Delta(\gamma), 0<\gamma<\pi/2,$ defined by  $ \delta_{kj} \le \gamma,$ we can take the gain $g=(1-\sin\gamma)(\pi/2-\gamma).$

For each transmission line $\{k,j\}$ connecting generator buses $k$ and $j,$ define the corresponding flow-in boundary segment 
$\partial\mathcal{P}_{kj}^{in}$ of the polytope $\mathcal{P}$ by equations/in-equations
$|\delta_{kj}|=\pi/2$ and $\delta_{kj}\dot{\delta}_{kj} < 0,$
and the flow-out boundary segment
$\partial\mathcal{P}_{kj}^{out}$  by
$|\delta_{kj}|=\pi/2$ and $\delta_{kj}\dot{\delta}_{kj} \ge 0.$
Consider the qudratic Lyapunov function $V(x)=x^TPx$ and define the following minimum value of the Lyapunov function $V(x)$ over the flow-out
boundary $\partial\mathcal{P}^{out}$ as:
\begin{align}\label{eq.Vmin}
 V_{\min}=\mathop {\min}\limits_{x \in \partial\mathcal{P}^{out}} V(x),
\end{align}
where $\partial\mathcal{P}^{out}$ is the union of
$\partial\mathcal{P}_{kj}^{out}$ over all the transmission lines $\{k,j\}\in \mathcal{E}$ connecting generator buses.  We have the following
result, which is a corollary of Theorem 1 in \cite{VuTuritsyn:2015TAC}. Hence, the proof is omitted.

\begin{theorem} 
\label{thr.StabilityAssessment}
  \emph{Consider a power system with the post-fault equilibrium point $\delta^* \in \Delta(\gamma)$ and the fault-cleared state $x_0$ staying in the polytope $\mathcal{P}.$ Assume that there exists a positive definite matrix $P$ such that
  \begin{align}
\label{eq.LMI}
\left[%
\begin{array}{cc}
 \bar{A}^TP+P\bar{A} +  \dfrac{(1-g)^2}{4}C^TC     & PB \\
 B^TP  & -I \\
\end{array}%
\right] \le 0 
\end{align}
and $
V(x_0) < V_{\min},
$
where $\bar{A}=A-\dfrac{1}{2}(1+g)BC.$
  Then, the system trajectory of \eqref{eq.structure-preserving} will converge from the fault-cleared state $x_0$ to the stable equilibrium point $\delta^*.$}
\end{theorem}

Therefore, a sufficient condition for the transient stability of the post-fault dynamics is the existence of a positive definite matrix $P$ satisfying the LMI \eqref{eq.LMI}
and the Lyapunov function at the fault-cleared state is small than the critical value $V_{\min}$ defined as in \eqref{eq.Vmin}. We will utilize this condition to design the emergency controls in the next sections.

\section{Fault-on Emergency Control Design}
\label{sec.FaultonControl}

\subsection{Control design}

In this section, we solve the problem $\textbf{(P1)}$, in which we maintain the power systems transient stability when a fault causes tripping of a line $\{u,v\}$. We will adjust the susceptances of some transmission lines during the fault-on dynamics. Applying Theorem 1, our objective is that: given a positive definite matrix $P$ satisfying the LMI \eqref{eq.LMI}, find the susceptances of the selected transmission lines such that the fault-cleared state $x_0$ satisfies $V(x_0)<V_{\min}.$

Note that the changing susceptances only affect the matrix $B,$ and then the fault-on dynamics with the tuned susceptances is described by
\begin{align}
\label{eq.Faulton}
\dot{x}_F=Ax_F-B(s)F(Cx_F) +B(s) D_{\{u,v\}}\sin\delta_{F_{uv}},
\end{align}
where $D_{\{u,v\}}$ is the vector to extract the $\{u,v\}$ element
from the vector of nonlinear interactions $F,$ while  $B(s)$ is the new system matrix $B$ obtained after the susceptances are changed. We have the following result the proof of which is in the Appendix \ref{app.ECdesign}.

\begin{theorem} 
\label{thr.ECdesign} \emph{Assume that there exist a positive definite
matrix $P$ of size $(|\mathcal{N}|+|\mathcal{G}|)$ satisfying the LMI \eqref{eq.LMI}. Let $\mu=\dfrac{\tau_{clearing}}{V_{\min}}$ 
where $V_{\min}$ is defined as in \eqref{eq.Vmin}. Assume that there exist feasible values for the susceptance of selected transmission lines and a positive definite matrix $\tilde{P}$  such that
\begin{align}
\label{eq.ECcondition2}
& \bar{A}(s)^T\tilde{P}+\tilde{P}\bar{A}(s) +  \dfrac{(1-g)^2}{4}C^TC   + \tilde{P}B(s)B(s)^T\tilde{P} \nonumber \\& +\mu \tilde{P}B(s)D_{\{u,v\}}D_{\{u,v\}}^TB(s)^T\tilde{P}  \le 0,
\end{align}
and 
\begin{align}
\label{eq.AdditionalCondition}
\tilde{P} \ge P
\end{align}
where $\bar A(s)=A-\dfrac{1}{2}(1+g)B(s)C.$
Then,  the fault-cleared state $x_0=x_F(\tau_{clearing})$ resulted from the fault-on dynamics \eqref{eq.Faulton}  is still inside the
region of attraction of the post-fault equilibrium point $\delta^*$, and the
post-fault dynamics following the tripping and reclosing of the line $\{u,v\}$ will return to
the original stable operating condition.}
\end{theorem}

Note that with fixed value of the susceptances, the inequality \eqref{eq.ECcondition2} can be rewritten by the following LMI with variable $\tilde{P}$:

\begin{align}
\label{eq.ECcondition3}
\left[%
\begin{array}{cc}
 \bar{A}(s)^T\tilde{P}+\tilde{P}\bar{A}(s) +  \dfrac{(1-g)^2}{4}C^TC     & \tilde{P}\bar{B}(s) \\
 \bar{B}(s)^T\tilde{P}  & -I \\
\end{array}%
\right] \le 0,
\end{align}
where $\bar{B}(s)=[B(s) \;\; \sqrt{\mu}B(s)D_{\{u,v\}}].$
Another way to solve the inequality  \eqref{eq.ECcondition2} is to take $\tilde{P}=P,$ and solve the LMI \eqref{eq.ECcondition3} with variable as susceptances of selected transmission lines.

\subsection{Procedure for Emergency Transmission Control on the Fault-on Dynamics}

We propose the following procedure to find suitable susceptance and execute fault-on emergency control:

\begin{itemize}
\item [1)] Find a positive definite matrix $P$ satisfying the LMI \eqref{eq.LMI}.
\item [2)] Calculate the minimum value $V_{\min}$ defined as in \eqref{eq.Vmin}.
\item [3)] Let $\mu=\dfrac{\tau_{clearing}}{V_{\min}}.$
\item [4)] Find the susceptances of selected transmission lines and  positive definite matrix $\tilde{P}$ satisfying the LMI  \eqref{eq.AdditionalCondition}-\eqref{eq.ECcondition3} by one of two ways described above.
\item [5)] If there is no such feasible values of susceptances and $\tilde{P}$, then repeat from step 1).
\item [6)] If such values of susceptance and positive definite matrix $\tilde{P}$ exist, then we keep these values
during the time period $[0,\tau_{clearing}].$ At the clearing time $\tau_{clearing},$ the fault is cleared and the susceptances of selected transmission lines are tuned back to their initial values.
\end{itemize}

\section{Post-Fault Emergency Control Design}
\label{sec.PostfaultControl}

In this section, we solve the post-fault emergency control $\textbf{(P2)}.$ Applying Theorem 1, to have a new stable post-fault dynamics with initial state $x_0$, the tuned values of susceptances need to satisfy three conditions:
\begin{itemize}
\item [(i)] There exists a new post-fault equilibrium point $\delta^*_{new}$  satisfying the power flow-like equations:
 \begin{align}
 \label{eq.PFE}
 \sum_{j\in \mathcal{N}_k}V_kV_jB_{kj}^{new}\sin \delta^*_{new_{kj}}=P_k, \forall k\in \mathcal{N}
 \end{align}
 
 \item [(ii)] There exists a positive definite matrix $P$ satisfying the LMI \eqref{eq.LMI} where $B=B(s)$ -the new matrix obtained after the susceptance is changed.
 
 \item [(iii)] The Lyapunov function $V(x)=x^TPx$ at the fault-cleared state $x_0$ (corresponding to $\delta_{fault-cleared}$) satisfies that $V(x_0)<V_{\min}$ where $V_{\min}$ is defined in \eqref{eq.Vmin}. 
\end{itemize}

We consider the special case when the fault-cleared state is a static point, i.e. $\dot\delta_{fault-cleared}=0$ and the fault-cleared state is only described by the angular $\delta_{fault-cleared_{kj}}.$ Instead of choosing the susceptance first and then solve the power flow-like equation \eqref{eq.PFE} to get the new equilibrium point, we use a heuristic procedure in which we select the new equilibrium point first and then find the susceptance from the power flow-like equation \eqref{eq.PFE}. Intuitively, to make sure that the fault-cleared state stays inside the region of attraction of the new equilibrium point, we need to select a desired equilibrium point as near the fault-cleared state as possible.  

\begin{figure}[t!]
\centering
\includegraphics[width = 3.2in]{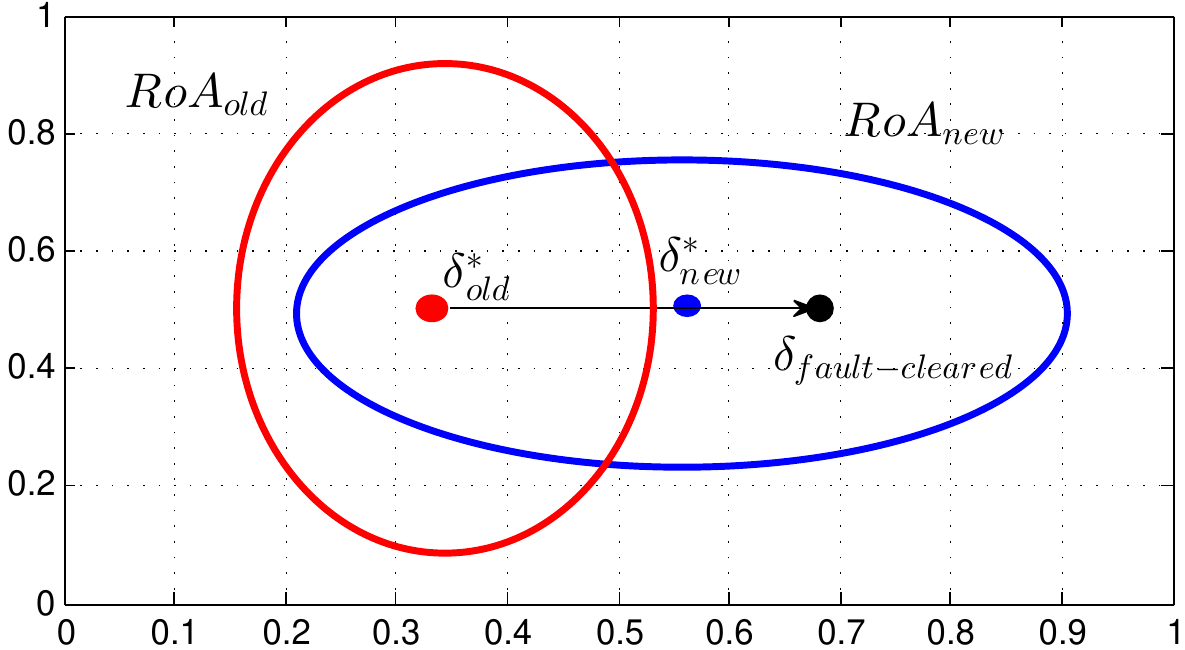}
\caption{Selection of the controlled post-fault equilibrium point $\delta^*_{new}$ such that its region of attraction contains the the fault-cleared state $\delta_{fault-cleared}$}
\label{fig.EmergencyControl_EquilibriumSelection}
\end{figure}

We propose the equilibrium selection as illustrated in Fig. \ref{fig.EmergencyControl_EquilibriumSelection},
where we select an equilibrium point between the old equilibrium point and the fault-cleared state such that this new equilibrium point is as near the fault-cleared state as possible, while satisfying the constraints on the possible changes of the susceptances. Note that if we allow the number of adjustable transmission lines larger than or equal to $n,$ then possibly we can always find the suitable susceptances satisfying the power flow-like equation \eqref{eq.PFE}. If the number of adjustable transmission lines smaller than $n,$ then we can use the convex optimizations to find the suitable susceptance such that $\sum_{j\in \mathcal{N}_k}V_kV_jB_{kj}^{new}\sin \delta^*_{fault-cleared_{kj}}$ is near $P_k$ (and therefore $\delta^*_{new}$ will be near $\delta^*_{fault-cleared}$). Indeed, let $y_k=P_k-\sum_{j\in \mathcal{N}_k}V_kV_jB_{kj}^{new}\sin \delta^*_{fault-cleared_{kj}}.$ Let the set of possible susceptance be $\mathcal{S}.$ Then we can have the optimization problem:
\begin{align}
\label{eq.SusceptanceSelection}
\min &  \sum_{k=1}^n y_k^2 \\
s.t. & \{B_{kj}\} \in \mathcal{S} \nonumber
\end{align}
When the acceptable set $\mathcal{S}$ is defined by linear constraints, by solving the convex optimization problem \eqref{eq.SusceptanceSelection} we can obtain the suitable susceptances such that the new equilibrium point is near $\delta_{fault-cleared}.$ From these suitable susceptances, we solve the power flow-like equation to get the new equilibrium point. After such new equilibrium point is found, we can find the positive definite matrix $P$ satisfying conditions (ii) and (iii) by using the adaptation algorithms presented in \cite{VuTuritsyn:2014} such that the stability region estimate corresponding with $P$ will contain the fault-cleared state $\delta_{fault-cleared}$.

\section{Numerical Validation}
\label{sec.simulations}

\subsection{Fault-on Emergency Control on 3 Generator System}
\begin{figure}[t!]
\centering
\includegraphics[width = 3.2in]{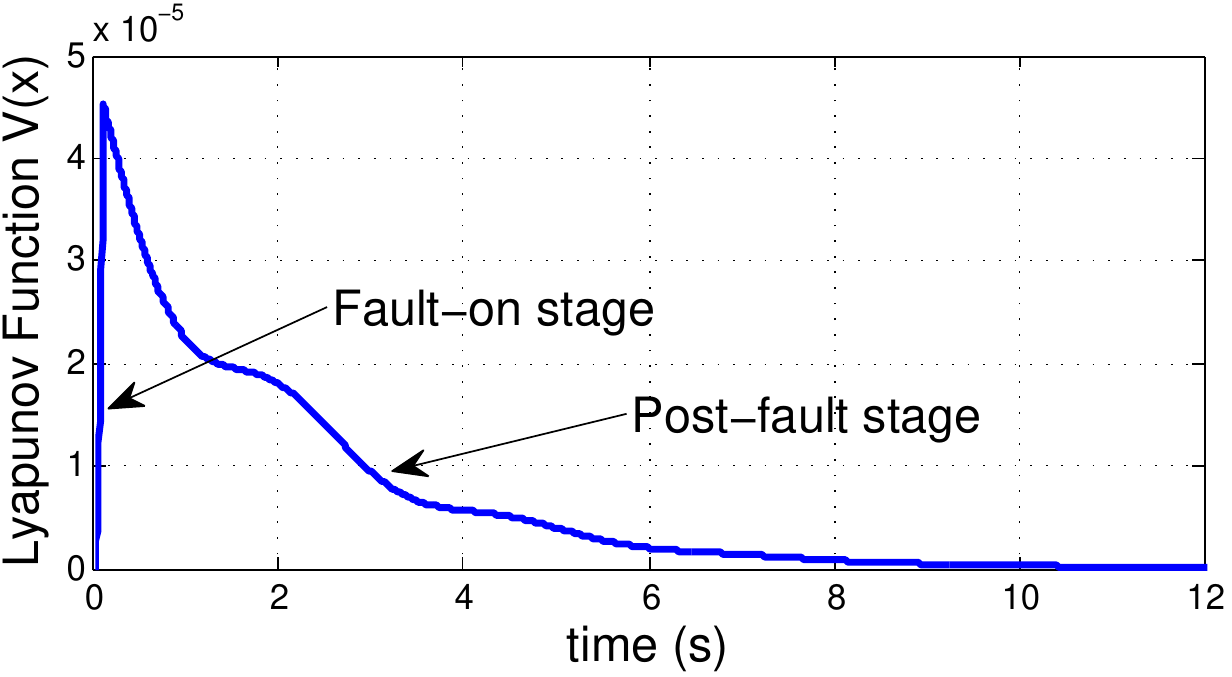}
\caption{Variations of the quadratic Lyapunov function $V(x)=x^TPx=(\delta-\delta^*)^TP(\delta-\delta^*)$ during the post-fault and fault-on dynamics}
\label{fig.FaultonEmergencyControl_Lyapunov}
\end{figure}

For illustrating the concept of this paper, we consider the simple yet non-trivial system of three generators, one of which is the renewable generator (generator 1) integrated with the synchronverter. 
The susceptance of the transmission lines are assumed at fixed values $B_{12}=0.739$ p.u., $B_{13}=1.0958$ p.u., and
$B_{23}=1.245$ p.u. Also, the inertia and damping of all the generators at the normal working condition are $m_k=2$ p.u., $d_k=1$ p.u. Assume that the line between generators 1 and 3 is tripped, and then reclosed at the clearing time $\tau_{clearing}=100ms,$ and during the fault-on dynamic stage the time-invariant terminal voltages and mechanical torques are $[V_1 \;V_2 \;V_3]=[1.0566 \;1.0502\;1.0170]$, $[P_1 \;P_2\; P_3]=[-0.2464\;0.2086\;0.0378]$.
The pre-fault and post-fault equilibrium point is calculated from \eqref{eq.SEP} as: $\delta^*=[-0.6634\;
   -0.5046\;
   -0.5640 \;0\;0\;0]^T.$ Hence, the equilibrium point stays in the polytope $|\delta_{kj}|<\pi/10.$ Then,  
   $g=(1-\sin(\pi/10))/(\pi/2-\pi/10).$ Using CVX to solve the LMI \eqref{eq.LMI}, we get the Lyapunov function with $P$ as
   \begin{align*}
  \left[%
\begin{array}{cccccc}
    2.8754  &  1.8587  &  1.9326  &  4.5165  &  4.3997  &  4.4179\\
    1.8587  &  2.8276  &  1.9805  &  4.4042  &  4.5088  &  4.4205\\
    1.9326  &  1.9805  &  2.7536  &  4.4127  &  4.4248  &  4.4950\\
    4.5165  &  4.4042  &  4.4127  & 18.0714  & 17.1319 &  17.2701\\
    4.3997  &  4.5088  &  4.4248  & 17.1319  & 17.9994 &  17.3418\\
    4.4179  &  4.4205  &  4.4950  & 17.2701 &  17.3418 &  17.8613\\
\end{array}%
\right]
   \end{align*}
Then the minimum value $V_{\min}$ is $V_{\min}= 0.9083$ and thus $\mu=\tau_{clearing}/V_{\min}= 0.1101.$ Assume that we can adjust the susceptance of transmission lines $\{1,2\}$ and $\{2,3\}$ within $50\%$ deviation from their initial value. 
Let $\tilde{P}=P,$ then we can solve the LMI \eqref{eq.ECcondition3}-\eqref{eq.AdditionalCondition} with variables $B_{12},B_{23}$, and obtain the optimum value of susceptances as $B_{12}^*= 0.7199 p.u., B_{23}^*=1.2093 p.u.$.  This means that there exist values of susceptances to compensate for the dynamics deviation caused by the faulted line ans thereby stabilize the power systems. This is confirmed in Fig. \ref{fig.FaultonEmergencyControl_Lyapunov}, where we can see the Lyapunov function $V(x)=x^TPx$ increases during the fault-on stage with feasible values of susceptances and then decreases to $0$ in the post-fault stage. 

\subsection{Post-Fault Emergency Control on 3 Generator System}

Now assume that we have a given initial state $\delta_{0}=\delta_{fault-cleared}=[0 \;0.5\;0.5 \;0 \;0\;0]^T$ and we want to stabilize the post-fault dynamics by adjusting the susceptances of the transmission lines 
$\{1,2\}$ and $\{2,3\}$.
Assume that the acceptable ranges for these susceptances are $0.4 \le B_{12}\le 1, 0.6 \le B_{23} \le 1.8$ (p.u.). Solving the convex optimization \eqref{eq.SusceptanceSelection} with the desired equilibrium point 
$\delta^*_{desired}=\delta_{fault-cleared},$ we obtain $B_{12}^*=0.4, B^*_{23}=1.2$ p.u. Using these new susceptances, we can calculate from the power flow-like equations the new equilibrium point as $\delta^*_{new}=[  -0.1403\;
    0.0766\;
   -0.0118 \;0\;0\;0]^T.$ Using the adaptation algorithm in \cite{VuTuritsyn:2014}, we can find the Lyapunov function $V(x)=x^TPx$ such that the stability region estimate corresponding with $P$ contains the fault-cleared state $\delta_{fault-cleared}$ 
   in which the matrix $P$ is 
   \begin{align*}
  \left[%
\begin{array}{cccccc}
    0.9881&   -0.0522&    0.0227&    0.7076&    0.5895&    0.6109\\
   -0.0522&    0.9520&    0.0588&    0.5962&    0.6979&    0.6161\\
    0.0227&    0.0588&    0.8771&    0.6132&    0.6296&    0.6901\\
    0.7076&    0.5962&    0.6132&   11.5526&   10.5799&   10.7195\\
    0.5895&    0.6979&    0.6296&   10.5799&   11.4685&   10.7746\\
    0.6109&    0.6161&    0.6901&   10.7195&   10.7746&   11.3194\\
\end{array}%
\right].
   \end{align*}

\section{Conclusions and path forward}
\label{sec.conclusion}

This paper proposed novel emergency control schemes for power grids by exploiting the plentiful transmission facilities. Particularly, we formulated two emergency control problems to maintain the transient stability of power systems: one involves the fault-on controlling to stabilize power systems following a given line tripping by intelligently and the other involves directly controlling the post-fault dynamics such that a given fault-cleared state stays inside the region of attraction of the new post-fault equilibrium point. In both problems, we applied our recently introduced quadratic Lyapunov function-based transient stability certificate \cite{VuTuritsyn:2015TAC} to give sufficient conditions for the adjusted susceptance of the selected transmission lines.  We showed that these problems can be solved through a number of convex optimizations in the form of linear matrix inequalities, which can be quickly solved by using advanced sparsity-exploiting SDP solvers \cite{Jabr2012}.

There are still many issues need to be addressed  to make these novel emergency control schemes ready for industrial employment. Particularly, we need to take into account the computation and regulation delays, either by offline scanning contingencies and calculating the emergency actions before hand, or by allowing specific delayed time for computation. Future works would demonstrate the proposed emergency control scheme on large IEEE prototypes and large dynamic realistic power systems with renewable generation at various locations and with different levels of renewable penetration. Also, a combination of the proposed method in this paper with the controlling UEP method \cite{Zou:2003ji,Chiangbook2015} promises to give us a less conservative, but simulation-free method for designing remedial actions from smart transmission facilities.

\section{Appendix}
\label{app.ECdesign}

Consider the Lyapunov function for the fault-on dynamics $\tilde{V}(x)=x^T\tilde{P}x$. Similar to the proof of Theorem 3 in \cite{VuTuritsyn:2015TAC}, From the inequality \eqref{eq.ECcondition2} we can show that
\begin{align}
\label{eq.Vfault}
\dot{\tilde{V}}(x_F) \le 1/\mu, \forall x_F \in \mathcal{P}
\end{align}

Now assume that $x_F(\tau_{clearing})$ is not in the set
$\mathcal{R}.$ Note that the boundary of $\mathcal{R}$ is constituted of the segments on flow-in boundary $\partial\mathcal{P}^{in}$ and the segments on the sublevel sets 
of the Lyapunov function. It is easy to see that the
flow-in boundary $\partial\mathcal{P}^{in}$ 
prevents the fault-on dynamics \eqref{eq.Faulton} from escaping $\mathcal{R}.$ Therefore, the fault-on trajectory can only escape
$\mathcal{R}$ through the segments which belong to sublevel set of
 $V(x).$ Denote $\tau$ be the first time at
which the fault-on trajectory meets one of the boundary segments
which belong to sublevel set of the Lyapunov function $V(x).$
Hence $x_F(t) \in \mathcal{R}, \forall 0 \le t \le \tau$ and $V(x_F(\tau))=V_{\min}.$ Noting
\eqref{eq.Vfault} and 
$\mathcal{R}\subset \mathcal{P},$ we have

\begin{align}
\label{eq.contradiction}
\tilde{V}(x_F(\tau))-\tilde{V}(x_F(0))  \le
\frac{\tau}{\mu} 
< \frac{\tau_{clearing}}{\mu}=V_{\min}
\end{align}
Note that $x_F(0)$ is the pre-fault equilibrium point, and thus equals to post-fault equilibrium point.
Hence $\tilde{V}(x_F(0))=0.$ Therefore, $\tilde{V}(x_F(\tau))<V_{\min}.$ Since $\tilde{P}\ge P,$ we have $V(x_F(\tau))\le \tilde{V}(x_F(\tau))<V_{\min},$
 a contradiction.


\bibliographystyle{IEEEtran}
\bibliography{lff}
\end{document}